\documentstyle [prl,aps,epsf]{revtex}
\begin{document}
\twocolumn[{
\widetext
\draft

\title{
Similarities between  organic and cuprate superconductors
}

\author{Ross H. McKenzie\cite{email}}

\date{{\it Science} {\bf 278}, 820 (1997).}
\maketitle

}]
\narrowtext

One of the greatest challenges of
condensed matter physics over the past
decade has been the theoretical 
description of the cuprate or high-temperature
superconductors.
When superconductivity was discovered in
organic solids   based on 
the BEDT-TTF molecule\cite{jerome} 
it was not clear if these materials were at all related
to the cuprates.
However, recent  work
has shown how the organics exhibit
some of the same interesting physics
as the cuprates including
unconventional metallic properties.

The family (BEDT-TTF)$_2$X consists of
conducting layers of BEDT-TTF molecules
sandwiched between insulating layers
of anions (X=Cu[N(CN)$_2$]Br, for example).
BEDT-TTF is a large planar molecule
and the different possible packing patterns
are denoted by different Greek letters \cite{jerome}.
The basic unit of the packing pattern
in the $\kappa$ phase is
a ``dimer'' consisting of two BEDT-TTF molecules
stacked on top of one another.
 Each dimer has one electron
less than a partially full electronic cloud
because of charge transfer to the anions.
The hole (missing electron) can hop
from dimer to dimer within a layer much easier than it can
hop between layers.
Consequently, as in the cuprates, the layered
structure leads to highly anisotropic 
electronic properties.
For example, the conductivity parallel to the layers is
two to five orders of magnitude larger than
perpendicular to the layers.

 The family $\kappa-$(BEDT-TTF)$_2$X
 has a particularly rich phase diagram as a function
of pressure, temperature, and anion, 
 as shown in the Figure \cite{kino,wzietek,kanoda}.
Note that
(i) antiferromagnetic and
 superconducting phases occur next to one another.
(ii) Recent experiments, to be discussed below,
show that the metallic phase has properties
that are quite distinct from conventional
metals.
(iii) The diagram is quite similar
to that of the cuprates if pressure is
replaced with doping.

Nuclear magnetic resonance (nmr) can be used to
probe the metallic state.
The electrons cause a shift in the resonant
frequency proportional to the electronic
density of states (Knight shift).
In a conventional metal, the Knight shift is independent
of temperature. In contrast, in
X=Cu[N(CN)$_2$]Br
it decreases significantly below about 50 K,
suggesting a suppression of the density of
states or ``pseudogap'' near the Fermi energy \cite{wzietek,desoto}.
If the nuclear spins are driven out of equilibrium 
then the rate  at which they realign with the external
field is determined by their interactions with the electrons.
The observed relaxation rate is five to ten
times larger than expected for a conventional metal
and strongly temperature dependent\cite{desoto}.
As the pressure is increased to four  kilobars the nmr 
properties become more like those of a conventional metal
\cite{wzietek}.
The temperature dependence of the nmr properties of
the cuprates
with small doping is similar\cite{ong}.

The unusual properties of the metallic state
are also seen in transport experiments.
As the temperature decreases to about 100K
the resistance increases, characteristic of a semiconductor.
Below 100 K it decreases rapidly and from
about 30 K down to the superconducting
transition temperature it decreases quadratically with
temperature. This temperature dependence is characteristic
of systems, such as transition
and heavy fermion metals, in which the dominant scattering mechanism is
the  interactions of the electrons with one
another. In  transition 
and heavy fermion metals there is a rule
relating the magnitude of the temperature
dependence to the density of states deduced 
from the electronic
specific heat. However, the observed scattering rate
in the organics is several hundred times larger than
predicted by this rule\cite{dressel}.
Furthermore, the scattering is so strong that above 30 K
the average distance an electron travels between collisions is
less than the lattice spacing. This
is inconsistent with conventional electronic transport.

In a conventional metal the Hall resistance
is a measure of the number of charge carriers
and is independent of temperature.
In contrast, 
recent measurements on the
X=Cu[N(CN)$_2$]Cl
 salt found a strong temperature dependence
\cite{sushko}.
 However, the ratio of the longitudinal
resistance to the Hall resistance has
a simple quadratic temperature dependence up to 100 K.
Similar behavior is found in the cuprates\cite{ong2}.

Experience with the cuprates
suggests the key physics
involved in the above behavior is
the layered structure  and strong interactions
between the electrons.
The importance of the latter is supported by
recent quantum chemistry calculations\cite{Fortu}
and by the large antiferromagnetic moment
observed in the insulating phase\cite{miyagawa}.

It should also be pointed out that
there is increasing evidence that, like in
the cuprates, 
the pairing of electrons in the superconducting     
state involves a different symmetry state 
than in conventional metals \cite{wzietek,kanoda}.

The theoretical challenge is to 
produce the simplest microscopic model
that can reproduce the phase diagram shown
in the Figure.  
Kino and Fukuyama\cite{kino}  recently
made some progress in this direction.
The key ingredients in their treatment
were that they took the  dimer 
structure   and the strong interactions
between electrons into account.
The role of increasing pressure is to 
reduce the amount of dimerization
and decrease the effect of the interactions
between the electrons.
Their model can describe the transition from the
insulating antiferromagnetic state to the metallic state 
and suggests the possibility of an intermediate
metallic antiferromagnetic state.
The limitations of their approach is that
it involves many parameters, only treats the magnetic
fluctuations in an average way, and does not predict
superconductivity.

The above interesting new findings show
that the organics are worthy of
more extensive study.
Theoretical studies should focus on
simplifying the           
model of Kino and Fukuyama
and should
take into account the magnetic fluctuations
 using techniques 
developed for the theory of the cuprate superconductors.
More experimental studies 
are needed to systematically 
characterise the unconventional properties of the
metallic state.\cite{review,support}

\begin{figure}
\centerline{\epsfxsize=9cm  \epsfbox{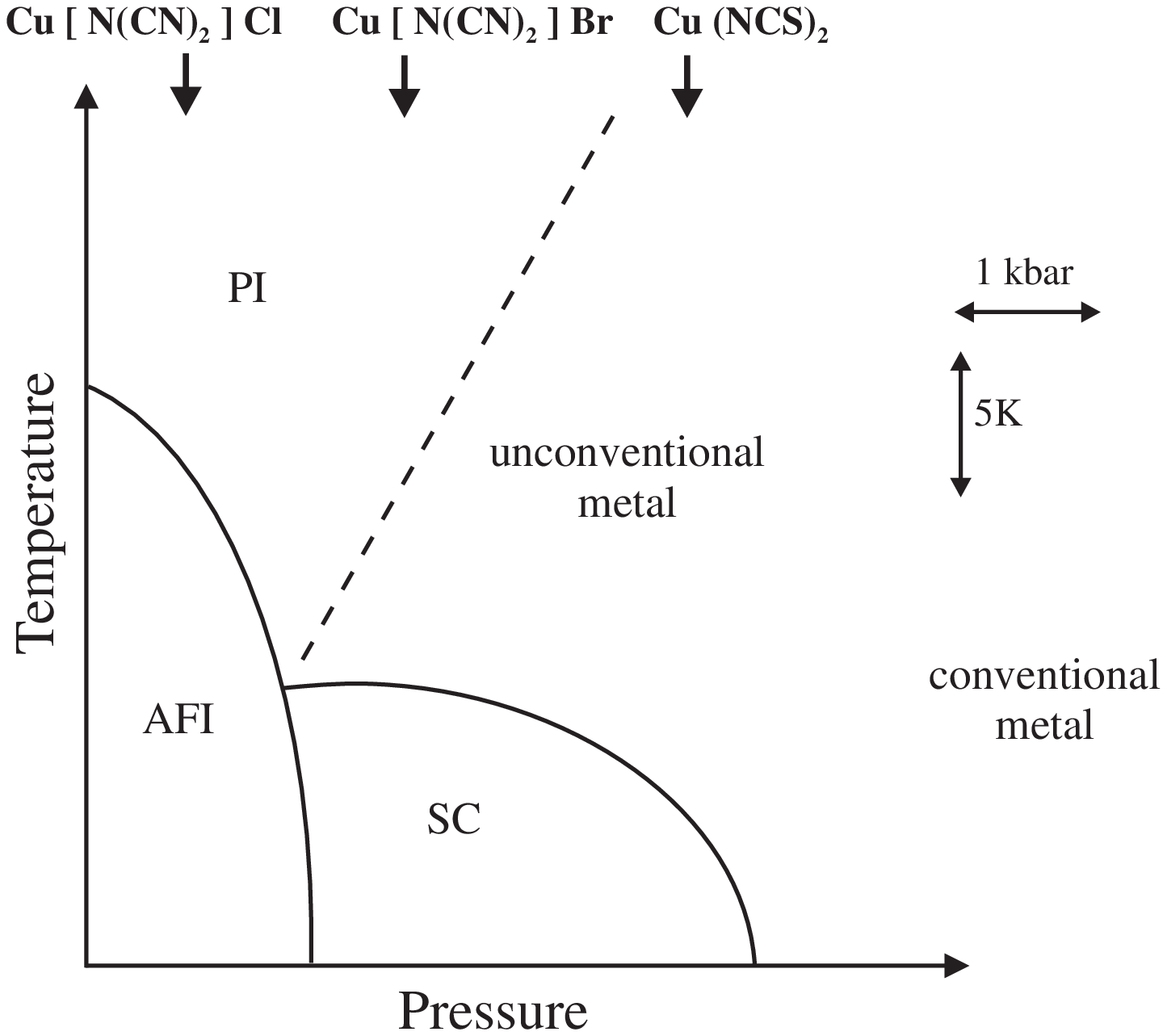}}
\vskip 0.5 cm
\caption{
Schematic phase diagram of the 
$\kappa-$(BEDT-TTF)$_2$X
family of organic conductors.
Superconducting, 
insulating antiferromagnetic and paramagnetic 
phases are denoted by SC, AFI, and PMI, respectively.
The arrows denote the location
of materials with different anions X
 at ambient pressure.
As the pressure decreases the 
properties of the metallic phase deviate
from those of a conventional metal.
The above phase diagram is qualitatively
similar to that of the cuprate superconductors
with doping playing the role of pressure.
\label{fig}} \end{figure}

\end{document}